\documentclass{jetpl}

\DeclareMathOperator{\sgn}{sgn}

\usepackage{graphicx}

\twocolumn
\lat

\begin{document}

 \title{Critical current in SFIFS junctions}
 \rtitle{Critical current in SFIFS junctions}
 \sodtitle{Critical current in SFIFS junctions}

 \author{A.\,A.~Golubov$^+$,\/\thanks{e-mail: a.golubov@tn.utwente.nl} M.\,Yu.~Kupriyanov$^*$, and
Ya.\,V.~Fominov$^{\ddag+}$}
 \rauthor{A.\,A.~Golubov, M.\,Yu.~Kupriyanov, and Ya.\,V.~Fominov}
 \sodauthor{Golubov, Kupriyanov, and Fominov}

\address{
$^+$ Department of Applied Physics, University of Twente, 7500 AE, Enschede, The Netherlands\\
$^*$ Nuclear Physics Institute, Moscow State University, 119899 Moscow, Russia\\
$^\ddag$ L.\,D.~Landau Institute for Theoretical Physics RAS, 117940 Moscow, Russia\\}

\dates{15 January 2002}{*}

\abstract{Quantitative theory of the Josephson effect in SFIFS junctions (S denotes bulk superconductor, F
--- metallic ferromagnet, I --- insulating barrier) is presented in the dirty limit. Fully self-consistent
numerical procedure is employed to solve the Usadel equations at arbitrary values of the F-layers
thicknesses, magnetizations, and interface parameters. In the case of antiparallel ferromagnets'
magnetizations the effect of the critical current $I_c$ enhancement by the exchange field $H$ is observed,
while in the case of parallel magnetizations the junction exhibits the transition to the $\pi$-state. In the
limit of thin F layers, we study these peculiarities of the critical current analytically and explain them
qualitatively; the scenario of the $0$--$\pi$ transition in our case differs from those studied before. The
effect of switching between $0$ and $\pi$ states by changing the F-layers' mutual orientation is
demonstrated.}


\maketitle

Josephson structures involving ferromagnets as weak link material are presently a subject of intensive
study. The possibility of the so-called ``$\pi$-state'' (characterized by the negative sign of the critical
current $I_c$) in SFS Josephson junctions was predicted theoretically
\cite{Bulaevsk1,Buzdin,Buzdin1,Buzdin2,Kriv,Radovic,Nazarov,Barash}. The first experimental observation of
the crossover from $0$- to $\pi$-state was reported by Ryazanov \textit{et al.} \cite{Ryazanov} and
explained in terms of temperature-dependent spatial oscillations of induced superconducting ordering in the
diffusive F layer.

More recently a number of new phenomena were predicted in junctions with more than one magnetically ordered
layer. First, the possibility of the critical current enhancement by the exchange field in SFIFS Josephson
junctions with thin F layers and antiparallel magnetization directions was discussed in the regimes of small
S layer thicknesses \cite{VolkovAF} and bulk S electrodes \cite{Koshina1,remark1}. Second, the crossover to
the $\pi$-state was predicted in Ref. \cite{Koshina1} for the parallel case even in the absence of the order
parameter oscillations in thin F layers. Still, the physical explanation of these effects and accurate
calculation of their magnitude have not been given so far. To make such estimates in the model with thin S
electrodes, one must consider KO-1 type solutions \cite{KO1} and take into account spatial variation of
superconducting state in the SF bilayers; at the same time, in the bulk S case an approximate method was
used in Ref. \cite{Koshina1} beyond its applicability range \cite{remark1}. This problem is of rather
general nature, since one may expect from the previous knowledge (see, e.g., review \cite{Likharev}) that
the supercurrent in a short weak link is $H$-independent.

The above intriguing scenario motivated us to attack the problem of the Josephson effect in SFIFS junctions
by self-consistent solution of the Usadel equations for arbitrary thicknesses of the F layers, barrier
transparencies and exchange field orientations. Below we show that the $0$--$\pi$ transition in the case of
parallel $H$ orientation or enhancement of $I_c$ by $H$ in the antiparallel case with thin F layers occurs
when the effective energy shift in the ferromagnets (due to the exchange field) becomes equal to a local
value of effective energy gap induced into a F layer. Under this condition a peak in the local density of
states (DoS) near the SF interfaces is shifted to zero energy. In the models with DoS of the BCS type this
leads to logarithmic divergency of $I_c$ in antiparallel case at zero temperature, similarly to the well
known Riedel singularity of \textit{ac} supercurrent in SIS tunnel junctions at voltage $eV=2\Delta$. We
also describe the general numerical method to solve the problem self-consistently and apply it for
quantitative description of the $0$--$\pi$ transition and $I_c$ enhancement in SFIFS junctions.

\textbf{The model.} We consider the structure of SFIFS type, where I is an insulating barrier of arbitrary
strength. We assume that the S layers are bulk and that the dirty limit conditions are fulfilled in the S
and F metals. Although our method is applicable in the general situation of different ferromagnets and
superconductors, for simplicity below we illustrate our results in the case when equivalent S and F
materials are used on both sides of the structure (although the directions of the exchange field in the two
F layers may be different), both F layers have the thickness $d_F$, and the two SF interfaces have the same
transparency. At the same time, we do not put any limitations on $d_F$ and the transparency.

The Usadel functions $G$, $F$ obey the normalization condition $G_\omega^2 + F_\omega F_{-\omega}^* =1$,
which allows the following parameterization in terms of the new function $\Phi$:
\begin{equation}
G_\omega= \frac{\widetilde\omega} {\sqrt{\widetilde\omega^2+\Phi_\omega \Phi_{-\omega}^*}},\qquad F_\omega
=\frac{\Phi_\omega} {\sqrt{\widetilde\omega^2+\Phi_\omega \Phi_{-\omega}^*}}. \label{def_f}
\end{equation}
The quantity $\widetilde\omega=\omega+iH$ corresponds to the general case when the exchange field $H$ is
present. However, in the S layers $H=0$ and we have simply $\widetilde\omega=\omega$.

We choose the $x$ axis perpendicular to the plane of the interfaces with the origin at the barrier I. The
Usadel equations \cite{Usadel} in the S and F layers have the form
\begin{gather}
 \xi_S^2 \frac{\pi T_c}{\omega G_S} \frac\partial{\partial x} \left[ G_S^2 \frac\partial
{\partial x} \Phi_S \right] -\Phi_S =-\Delta,  \label{EqUS}\\
 \xi_F^2 \frac{\pi T_c}{\widetilde\omega G_F} \frac\partial{\partial x} \left[ G_F^2 \frac\partial
{\partial x} \Phi_F \right] -\Phi_F =0, \label{EqUSF}
\end{gather}
where $T_c$ is the critical temperature of the superconductors, $\Delta$ is the pair potential (which is
nonzero only in the S layers), $\omega$ is the Matsubara frequency, and the coherence lengthes $\xi$ are
related to the diffusion constants $D$ as $\xi_{S(F)}=\sqrt{D_{S(F)}/2\pi T_c}$. The pair potential
satisfies the self-consistency equations
\begin{equation}
\Delta \ln \frac T{T_c} +\pi T\sum_\omega \frac{\Delta -G_S \Phi_S \sgn\omega}{|\omega|} =0. \label{EqDEL}
\end{equation}
In the present paper we restrict ourselves to the cases of parallel and antiparallel orientations of the
exchange fields $H$ in the ferromagnets.

The boundary conditions at the SF interfaces ($x=\mp d_F$) have the form \cite{KL} (see Ref. \cite{Koshina2}
for details)
\begin{gather}
\frac{\xi_S G_S^2}\omega \frac\partial{\partial x} \Phi_S =\gamma \frac{\xi_F G_F^2} {\widetilde\omega}
\frac\partial{\partial x}
\Phi_F, \label{BC_Fi2} \\
\pm\gamma_B \frac{\xi_F G_F}{\widetilde\omega} \frac\partial{\partial x} \Phi_F = G_S \left(
\frac{\Phi_F}{\widetilde\omega} -\frac{\Phi_S}\omega \right), \label{BC_Fi1} \\
\text{with}\quad \gamma_B= R_B \mathcal{A} / \rho_F \xi_F,\quad \gamma= \rho_S \xi_S / \rho_F \xi_F, \notag
\end{gather}
where $R_B$ and $\mathcal{A}$ are the resistance and the area of the SF interfaces; $\rho_{S(F)}$ is the
resistivity of the S (F) layer. At the I interface ($x=0$) the boundary conditions read
\begin{gather}
\frac{G_{F1}^2}{\widetilde\omega_1} \frac\partial{\partial x} \Phi_{F1} =
\frac{G_{F2}^2}{\widetilde\omega_2} \frac\partial{\partial x} \Phi_{F2}, \label{BI_1} \\
\gamma_{B,I} \frac{\xi_F G_{F1}}{\widetilde\omega_1} \frac\partial{\partial x} \Phi_{F1} = G_{F2} \left(
\frac{\Phi_{F2}}{\widetilde\omega_2} -\frac{\Phi_{F1}}{\widetilde\omega_1} \right), \label{BI_2} \\
\text{with}\quad \gamma_{B,I}= R_{B,I} \mathcal{A} / \rho_F \xi_F, \notag
\end{gather}
where the indices $1,2$ refer to the left and right hand side of the I interface, respectively.

In the bulk of the S electrodes we assume a uniform current-carrying superconducting state
\begin{equation}
\Phi(x=\mp \infty) =\frac{\Delta_0 \exp\left(i[\mp \varphi /2+ 2m v_s x]\right)}{1+2D_S m^2 v_s^2
/\sqrt{\omega^2+|\Phi|^2}}, \label{BC_bulk}
\end{equation}
where $m$ is the electron's mass, $v_s$ is the superfluid velocity, and $\varphi$ is the phase difference
across the junction.

The supercurrent density is constant across the system. In the F part it is given by the expression
\begin{equation}
J=\frac{i\pi T}{2e\rho}\sum_\omega \frac{G^2(\omega)}{\widetilde\omega^2}\left[ \Phi_\omega
\frac\partial{\partial x} \Phi_{-\omega}^* -\Phi_{-\omega}^* \frac\partial{\partial x} \Phi_\omega \right],
\label{cur}
\end{equation}
while analogous formula for the S part is obtained if we substitute $\widetilde\omega\to \omega$. This
expression, together with the boundary condition (\ref{BI_2}) and the symmetry relation
$F(-\omega,H)=F(\omega,-H)$, yields the formula for the supercurrent across the I interface:
\begin{equation} \label{cur1}
I = \frac{\pi T}{e R_{B,I}} \sum_\omega \Imag\left[ F_{F1}^*(-H_1) F_{F2}(H_2) \right]
\end{equation}
[the functions $F$ are related to $\Phi$ via Eq. (\ref{def_f})].

\textbf{The limit of small F-layer thickness: $d_F \ll \min(\xi_F,\sqrt{D_F/2H})$.} Under the condition
$\gamma_B/\gamma\gg 1$ we can neglect the suppression of superconductivity in the superconductors. We assume
further that the transparency of the barrier I is small, $\gamma_{B,I}\gg\max(1,\gamma_B)$, and the SF
bilayers are decoupled (the exact criterion will be given below). In this case we can set $v_s=0$ and expand
the solution of Eq. (\ref{EqUSF}) in the F layers up to the second order in small spatial gradients.
Applying the boundary condition (\ref{BC_Fi1}), we obtain the solution in the form similar to that in SN
bilayer \cite{Gol1,Koshina2}:
\begin{gather}
\Phi_{F1,F2} = \frac{\widetilde\omega_{1,2} /\omega}{1+\gamma_{BM} \widetilde\omega_{1,2} / \pi T_c G_S}
\Delta_0 \exp(\mp i\varphi/2),  \label{Sol2}\\
\text{with}\quad \gamma_{BM}=\gamma_{B} d_F /\xi_F,\quad G_S=\omega/\sqrt{\omega^2+\Delta_0^2}.\notag
\end{gather}
Substituting Eq. (\ref{Sol2}) into the expression for the supercurrent (\ref{cur1}) we obtain
$I(\varphi)=I_c \sin\varphi$.

For the parallel orientation of the exchange fields, $H_1=H_2=H$, the critical current is
\begin{equation}
I_c^{(p)}=\frac{2\pi T}{eR_{B,I}}\sum_{\Omega >0} \frac{\delta^2 G_S^2}{\Omega^2} \frac{1-\alpha +\Omega
\gamma_{BM} g_1}{(1-\alpha +\Omega \gamma_{BM}g_1)^2+4\alpha g_2}, \label{CurPF}
\end{equation}
where $\Omega =\omega /\pi T_c$, $\delta =\Delta_0 /\pi T_c$, $\alpha =(h\gamma_{BM})^2$, $h=H/\pi T_c$,
$g_1= 2 G_S+\gamma_{BM}\Omega$, $g_2=(G_S+\gamma_{BM} \Omega)^2$.

For the antiparallel orientation, $H_1=-H_2=H$, the critical current is given by
\begin{equation}
I_c^{(a)}=\frac{2\pi T}{eR_{B,I}} \sum_{\Omega >0} \frac{\delta^2 G_S^2}{\Omega^2} \frac 1{\sqrt{(1-\alpha
+\Omega \gamma_{BM} g_1)^2+4\alpha g_2}}.  \label{CurAF}
\end{equation}
At $h=1/\gamma_{BM}$ and small $\Omega$ the expression under the sum in Eq. (\ref{CurAF}) behaves as
$1/\Omega$, thus at low $T$ the critical current diverges logarithmically: $I_c^{(a)} \propto \ln(T_c/T)$.
This effect was pointed out earlier in Refs. \cite{VolkovAF,Koshina1}.

The above results become physically transparent in the real energy $\varepsilon$ representation. Making
analytical continuation in Eqs. (\ref{def_f}), (\ref{Sol2}) by replacement $\omega \rightarrow
-i\varepsilon$, we obtain the expression for the DoS per one spin projection (spin ``up'')
$N_F(\varepsilon)=\Real G_F(\varepsilon)$ in the F layers
\begin{gather}
N_F(\varepsilon)= \left| \Real \frac{\widetilde\varepsilon}{\sqrt{\widetilde\varepsilon^2-\Delta_0^2}}
\right|, \label{DENS}\\
\widetilde\varepsilon =\varepsilon +\gamma_{BM}(\varepsilon -H)\sqrt{\Delta_0^2-\varepsilon^2}/ \pi
T_c,\notag
\end{gather}
which demonstrates the energy renormalization due to the exchange field. Equation (\ref{DENS}) yields
$N_F(0)=\Real ( \gamma_{BM}h /\sqrt{(\gamma_{BM} h)^2-1})$, which shows that at $h=1/\gamma_{BM}$ the
singularity in the DoS is shifted to the Fermi level. Exactly at this value of $h$ the maximum of
$I_c^{(a)}$ is achieved due to overlap of two $\varepsilon^{-1/2}$ singularities. This leads to logarithmic
divergency of the critical current (\ref{CurAF}) in the limit $T\to 0$, similarly to the well known Riedel
singularity of nonstationary supercurrent in SIS tunnel junctions at voltage $eV=2\Delta_0$, where the
energy shift is due to the electric potential. At the same value of the exchange field $h=1/\gamma_{BM}$ the
critical current changes its sign (i.e., the crossover from $0$ to $\pi$ contact occurs) for parallel
magnetizations in the F layers [see Eq. (\ref{CurPF})]. We emphasize that the scenario of the $0$--$\pi$
transition in our case differs from those studied before where the $\pi$-shift of the phase was either due
to spatial oscillations of the order parameter in F layers or due to the proximity-induced phase rotation in
S layers. In our case the phase does not change in either layer; instead, it jumps at the SF interfaces.
This scenario is most clearly illustrated in the limit of large $H$ where Eqs. (\ref{def_f}), (\ref{Sol2})
yield $F_F \propto -i\Delta\sgn H $ whereas $F_S \propto \Delta$; thus the phase jumps by $\pi/2$ at each of
the SF interfaces, providing the total $\pi$-shift between $F_{F1}(-H)$ and $F_{F2}(H)$ [it is the phase
difference between these two functions that determines the supercurrent according to Eq. (\ref{cur1})].

The considered effects take place only for sufficiently low I-barrier transparency. Indeed, it follows from
Eq. (\ref{Sol2}) that $G_F (\Omega)\propto 1/\sqrt{\Omega}$ for small $\Omega$ under condition
$h=1/\gamma_{BM}$. As a result, the boundary condition (\ref{BI_2}) yields that at
\begin{equation}
\Omega \leq \min \left( \frac {\xi_F} {d_F\gamma_{B,I}},\; \frac{\gamma_B}{\gamma_{B,I}} \right)
\label{condT}
\end{equation}
the solutions (\ref{Sol2}) are not valid, since in this frequency range the effective transparency of the I
interface (the parameter $G_{F1} G_{F2} / \gamma_{B,I}$ \cite{we}) increases and the spatial gradients in
the F layers become large (the limit of large gradients is called ``the KO-1 case'' \cite{KO1,Likharev}). In
this case the nongradient term in Eq. (\ref{EqUSF}) can be neglected and the general solution of the Usadel
equation in the F layers has the KO-1 form \cite{KO1}:
\begin{equation} \label{Sol1}
\frac\Phi{\widetilde\omega} = \frac{ C-i M\arctan \left[ M\left( B x+Q \right) \right]}{1-\eta} ,
\end{equation}
where $M=\sqrt{(\eta^2-1)-C^2}$, while $C$, $B$, $Q$ and $\eta$ are integration constants. From Eqs.
(\ref{def_f}), (\ref{Sol1}) it follows that the Green functions $G$, $F$ and hence the contribution to the
critical current from these frequencies are $H$-independent. As a result, the barrier transparency parameter
$\gamma_{B,I}$ provides the cutoff of the low-temperature logarithmic singularity of $I_c^{(a)}$ at
$h=1/\gamma_{BM}$ [see Eq. (\ref{CurAF})]. According to Eq. (\ref{condT}), the critical current saturates at
low temperature $T^*=T_c\min ( \xi_F/d_F\gamma_{B,I},\; \gamma_B/\gamma_{B,I} )$. We note that any asymmetry
in the SFIFS junction will also lead to the cutoff of $I_c^{(a)}$ divergency \cite{we}. The above estimates
are done for the case of low barrier transparency, $\xi_F/d_F\gamma_{B,I}\ll1$ and
$\gamma_B/\gamma_{B,I}\ll1$. The opposite regime of high transparency deserves separate study.

\textbf{The general case.} For arbitrary F-layer thicknesses and interface parameters the boundary problem
(\ref{def_f})--(\ref{BC_bulk}) has been solved numerically using iterative procedure. Starting from trial
values of the complex pair potentials $\Delta$ and the Green functions $G_{S,F}$ we solve the resulting
linear equations and boundary conditions for functions $\Phi_{S,F}$. After that we recalculate $G_{S,F}$ and
$\Delta$. Then we repeat the iterations until convergency is reached. The self-consistency of calculations
is checked by the condition of conservation of the supercurrent (\ref{cur}) across the junction. We
emphasize that our method is \textit{fully} self-consistent: in particular, it includes the self-consistency
over the superfluid velocity $v_s$, which is essential (contrary to the constriction case) in the
quasi-one-dimensional geometry. The details of our numerical method will be presented elsewhere \cite{we}.

\begin{figure}
 \centerline{\includegraphics[width=84mm]{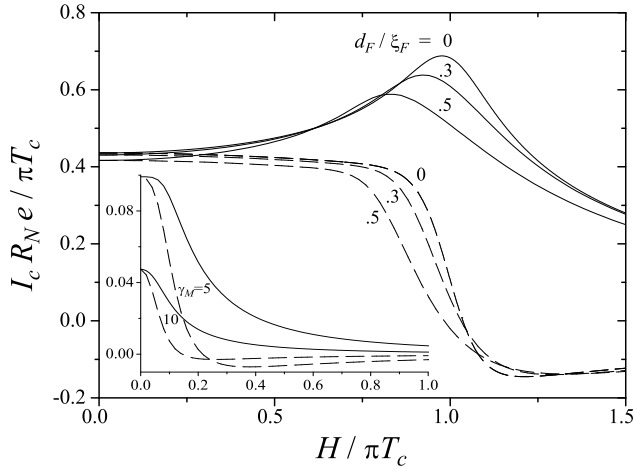}}
 \caption{Fig.\ref{fig:fig1}. Enhancement of the critical current (antiparallel magnetizations, solid
 lines) and the $0$--$\pi$ transition at which $I_c$ changes its sign (parallel magnetizations, dashed
 lines) in the SFIFS junction at
 $T/T_c = 0.05$, $\gamma_{BM}=1$, and $\gamma_M=0$. Inset: the same for large values of $\gamma_M$.}
 \label{fig:fig1}
\end{figure}

Figure~\ref{fig:fig1} shows $I_c(H)$ dependencies calculated at $T=0.05\, T_c$ from the numerical solution
of the boundary problem (\ref{def_f})--(\ref{BC_bulk}) for the fixed value of $\gamma_{BM}=1$ and a set of
different F-layers thicknesses and the SF interface parameters $\gamma$. The normal junction resistance is
$R_N=R_{B,I}+2R_B+ 2\rho_F d_F/ \mathcal{A}$. The curves $d_F/\xi_F=0$ are the limits of vanishing
$d_F/\xi_F$ ratio at fixed $\gamma_{BM}$ and are calculated from Eqs. (\ref{CurPF}), (\ref{CurAF}). For thin
F layers the results depend only on the combination $\gamma_M=\gamma d_F/\xi_F$. The enhancement of $I_c$
and the crossover to the $\pi$-state are clearly seen for the antiparallel and parallel orientations,
respectively. In accordance with the estimates given above, these effects take place for the values of the
exchange field $H$ close to $\pi T_c$. The enhancement disappears with increasing gradients in the F layers
since the solution Eq. (\ref{Sol2}) loses its validity. This is illustrated in Fig.\ref{fig:fig1} by
increasing the thickness $d_F$ or $\gamma_M$. In particular, in the case of large $\gamma_M$ the enhancement
is absent, in contrast to the statement in Ref. \cite{Koshina1} (see \cite{remark1}).

\begin{figure}
 \centerline{\includegraphics[width=84mm]{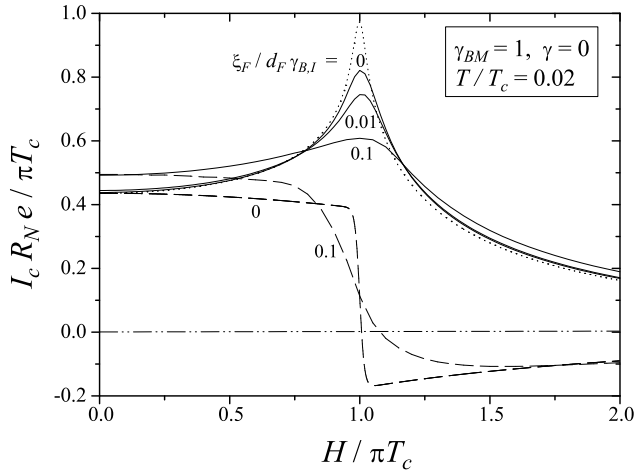}}
 \caption{Fig.\ref{fig:fig2}. Enhancement of the critical current (antiparallel magnetizations, solid
 lines) and the $0$--$\pi$ transition at which $I_c$ changes its sign (parallel magnetizations, dashed
 lines) in the SFIFS junction:
 influence of temperature and barrier transparency. The dotted line corresponds to $T/T_c=0.01$ and
 $\xi_F / d_F \gamma_{B,I}=0$; the parameters for other curves are given in the Figure.}
 \label{fig:fig2}
\end{figure}

Influence of temperature and barrier transparency on the critical current anomaly is shown in
Fig.\ref{fig:fig2}. One can see that, in accordance with the above estimate, the cutoff of $I_c^{(a)}$
singularity is provided by finite temperature or barrier transparency. Namely, with the decrease of the
barrier strength parameter $\gamma_{B,I}$ the peak magnitude starts to drop when the ratio
$d_F\gamma_{B,I}/\xi_F$ becomes comparable to $T/T_c$. With further decrease of $d_F\gamma_{B,I}/\xi_F$ the
singularity disappears, while the transition to the $\pi$-state shifts to large values of $H$.

\begin{figure}
 \centerline{\includegraphics[width=84mm]{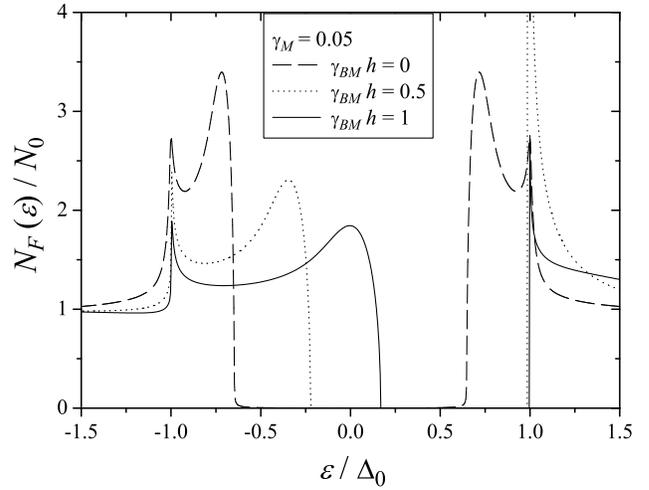}}
 \caption{Fig.\ref{fig:fig3}. Normalized density of states for spin ``up'' in the F layer for various
 exchange fields.}
 \label{fig:fig3}
\end{figure}

Figure~\ref{fig:fig3} demonstrates the DoS in the F layers for one spin projection, calculated numerically
in the limit of small I-barrier transparency. At $H=0$ we reproduce the well-known minigap existing in SN
bilayer. At finite $H$ the gap shifts in energy (asymmetrically) and the peak in the DoS reaches zero energy
at $h=1/\gamma_{BM}$. One can see that even for a small value $\gamma_M = 0.05$ the peaks are rather broad,
this is the reason why the singularity in $I_c^{(a)}$ is suppressed by $\gamma_M$ very rapidly.

\begin{figure}
 \centerline{\includegraphics[width=84mm]{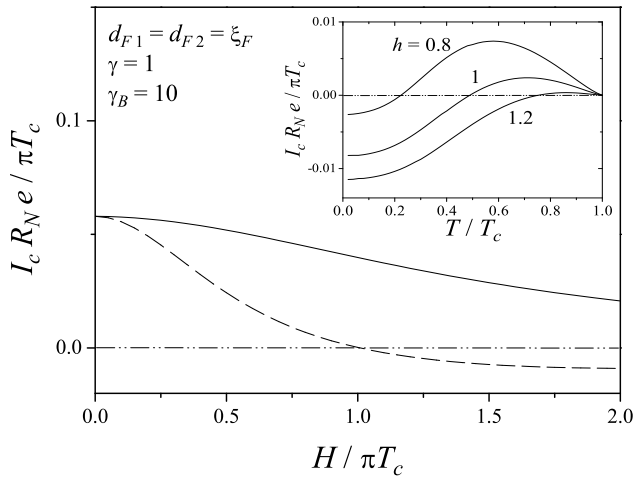}}
\caption{Fig.\ref{fig:fig4}. Critical current in SF$_1$F$_2$S junction: switching effect. $T/T_c=0.5$, the
solid and dashed lines correspond to the antiparallel and parallel orientations of magnetizations,
respectively. Inset: thermally induced $0$--$\pi$ crossover in the parallel case.}
 \label{fig:fig4}
\end{figure}

In the practically interesting limit of finite F-layer thickness (see Fig.\ref{fig:fig4}) the numerical
calculations show monotonic suppression of $I_c$ with increase of the exchange field $H$ for antiparallel
magnetizations of the F layers and the $0$--$\pi$ crossover for the parallel case. One can see from
Fig.\ref{fig:fig4} that for given temperature and thickness of the F layers it is possible to find the value
of the exchange field at which switching between parallel and antiparallel orientations will lead to
switching of $I_c$ from nearly zero to a finite value (or to switching between $0$ and $\pi$ states). This
effect may be used for engineering cryoelectronic devices manipulating spin-polarized electrons.

The case of parallel F-layers magnetizations in the absence of the I barrier corresponds to the standard SFS
junction where the $0$--$\pi$ transition is possible due to spatial oscillations of induced superconducting
ordering in the F layer. The thermally induced $0$--$\pi$ crossover in SFS junction was observed in Ref.
\cite{Ryazanov}, where simple theory based on the linearized Usadel equations was also presented. Here we
show such a crossover (see the inset in Fig.\ref{fig:fig4}) from the fully self-consistent solution in the
range of the exchange fields corresponding to that of Ref. \cite{Ryazanov}. Comparison to the experimental
data and more detailed results of our model will be given elsewhere \cite{we}.

In conclusion, we have presented a general method to solve the Usadel equations in SFIFS junctions
self-consistently. Using our method, we have investigated theoretically the Josephson current in SFIFS and
SFS junctions as a function of relative F-layers magnetizations, thicknesses and parameters of the S/F and
F/F interfaces. We have identified the physical mechanisms of the critical current enhancement and of the
$0$--$\pi$ transition in these junctions.

We acknowledge stimulating discussions with J.~Aarts, N.\,M.~Chtchelkatchev, K.\,B.~Efetov,
M.\,V.~Feigel'man, V.\,V.~Ryazanov, and M.~Siegel. The research of M.Yu.K. was supported by the Russian
Ministry for Industry and Technology. Ya.V.F. acknowledges financial support from the Russian Foundation for
Basic Research (project 01-02-17759), and from Forschungszentrum J\"ulich (Landau Scholarship).

\end{document}